\documentclass[twoside]{dis08}
\usepackage[latin1]{inputenc}
\usepackage[dvips]{graphicx,epsfig,color}
\usepackage{wrapfig,rotating}
\usepackage{amssymb,amsmath,array}

\begin{document}
\title{\raggedleft{\normalsize DESY 08-110} \\[0.5em]
Diffraction and Forward Physics: from HERA to LHC%
\,\thanks{Talk given at the 16th International Workshop on Deep Inelastic
    Scattering and Related Subjects, University College London, England,
    7--11 April 2008 \protect\cite{url}}
}
\author{Markus Diehl
\vspace{.3cm}\\
Deutsches Elektronen-Synchroton DESY \\
22603 Hamburg - Germany
}

\maketitle

\begin{abstract}
  I discuss and connect a number of topics in small-$x$ physics at HERA
  and at LHC, pointing out recent progress and open questions in theory
  and phenomenology.
\end{abstract}


\section{Leading protons and rapidity gaps}

An anticipated highlight of diffraction at LHC is the study of new
particles in central exclusive production, $pp\to p+X+p$.  Detailed
investigations have been made for the case where $X$ is a Higgs boson in
the Standard Model or its supersymmetric extension (see e.g.\
\cite{Heinemeyer:2007tu}), but other systems with a strong coupling to two
gluons, like a gluino pair $X=\tilde{g}\tilde{g}$ \cite{Bussey:2006vx},
can be equally interesting.  Provided that rates are sufficiently high,
central exclusive production enables us to study the system $X$ in a clean
environment, with a signal-to-background ratio often much larger than in
conventional, inclusive production channels.  Measurement of the outgoing
proton momenta in forward detectors gives the possibility of a precise
determination of the mass and possibly the width of $X$, and the exclusive
production mechanism strongly favours systems $X$ with quantum numbers
$CP=++$.  If the effective two-gluon luminosity for $pp\to p+gg+p \to
p+X+p$ can be determined from Standard-Model channels such as
$X=\text{dijet}$ or $X=\gamma\gamma$, the cross section measurement for a
new particle $X$ decaying into a final state $f$ yields the combination
$\Gamma_{X\to gg} \Gamma_{X\to f}/ \Gamma_{\text{tot}}$ of widths.  More
detail is given in several presentations at this meeting \cite{CEP-talks}.

\begin{wrapfigure}{r}{0.5\columnwidth}
\centerline{\includegraphics[width=0.371\columnwidth]{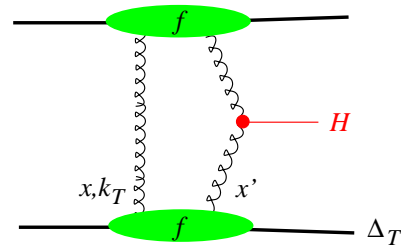}}
\caption{\label{fig:exc-higgs} A simple graph for central exclusive Higgs
  production.  The blobs represent the generalised gluon distribution.}
\end{wrapfigure}
HERA data provides essential non-perturbative input needed to calculate
central exclusive production at LHC.  This is important not only to
estimate event rates before LHC measurements start, but also to control
backgrounds and to help optimise triggers and selection cuts when LHC data
will be available.  An example are events where the central system $X$
contains not only a Higgs but also relatively soft gluons.  Diffractive
parton densities from HERA \cite{{DPDF-talks}} are crucial to estimate the
importance of this channel, which would spoil the accuracy of a Higgs mass
determination from the scattered protons alone \cite{Khoze:2007hx}.

The theory description of central exclusive production involves a number
of nontrivial issues, some of which were discussed at this meeting
\cite{Khoze-talk}.  The simplest leading-order graph for the production
mechanism is shown in Fig.~\ref{fig:exc-higgs}.  There is no all-order
factorisation theorem for this process, and indeed factorisation is broken
by rescattering between spectator partons.  Radiative corrections to the
leading-order graph in Fig.~\ref{fig:exc-higgs} are known to be important
and can in part be grouped into Sudakov factors.  A full next-to-leading
order calculation is, however, not available, and it has recently been
argued that there may be important higher-order corrections which have not
been evaluated so far \cite{Cudell:2008gv}.

The non-perturbative input to the graph in Fig.~\ref{fig:exc-higgs} is
given by the generalised gluon distribution.  Our present knowledge of
this distribution derives essentially from HERA data on vector meson
production and on deeply virtual Compton scattering \cite{VM-DVCS-talks}.
From a theory point of view, $\Upsilon$ photoproduction is a particularly
clean channel for this purpose, but the rate at HERA is too limited for
detailed measurements.  Along with $J\!/\Psi$ photoproduction this channel
could, however, be further studied in $pp$ and $pA$ collisions at LHC
\cite{LHC-gamma-talks}.  It is important to note that the calculation of
central exclusive production involves generalised distributions $f^g$ that
depend explicitly on the transverse momentum $k_T$ of the emitted gluon.
Put in a simplified way, the graph in Fig.~\ref{fig:exc-higgs} involves an
integral
\begin{equation}
\int \frac{d^2 k_T}{k_T^4}\, f^g(x_1^{},x_1',\Delta_{1T}^{},k_T^{})\,
                             f^g(x_2^{},x_2',\Delta_{2T}^{},k_T^{}) \,,
\end{equation}
where the distributions in the two colliding protons are entangled.  This
cannot readily be reduced to the $k_T$ integrated, collinear distributions
\begin{equation}
\int^{\mu^2} \frac{d k_T^2}{k_T^2}\, f^g(x,x',\Delta_{T},k_T)
\end{equation}
that appear in $ep$ scattering processes.  While the understanding and
evaluation of higher-order corrections is fairly advanced for collinear
generalised parton distributions, the same is unfortunately not the case
for their $k_T$ dependent counterparts.

As already mentioned, the simple mechanism shown in
Fig.~\ref{fig:exc-higgs} receives important corrections from the
rescattering of spectator partons in the two colliding protons.  Most
calculations assume that the dominant rescattering effects can be
described by elastic or quasi-elastic proton-proton interactions.  This
leads to a simple representation
$\sigma = \sigma_{\text{hard}} \otimes |S|^2$
for the physical cross section, where the convolution is in
transverse-momentum space, $\sigma_{\text{hard}}$ describes the
hard-scattering mechanism sketched in Fig.~\ref{fig:exc-higgs}, and the
rapidity gap survival factor $|S|^2$ can be inferred from $pp$ and
$p\bar{p}$ scattering data.  More complicated rescattering mechanisms
have, however, been studied \cite{Bartels:2006ea,Miller-talk}.  The
underlying physics is related to multiple scattering in inclusive $pp$
collisions, which by itself is of high importance for understanding final
states at LHC.  The description of such rescattering effects involves
considerable difficulties, and a reliable evaluation of their importance
has not been achieved as yet.

It is all the more important to test the phenomenological models currently
used for the description of central exclusive production.  Fortunately,
this is already possible by using data from the Tevatron, either for
inclusive diffractive channels such as $\bar{p}p \to \bar{p}+ \text{dijet}
+X$ or for exclusive reactions like $\bar{p}p \to \bar{p}+ \text{dijet}
+p$ and $\bar{p}p \to \bar{p}+ \gamma\gamma+ p$ \cite{Tev-talks}.
Possible tests using early data from LHC are discussed in
\cite{Khoze-talk}.  HERA provides crucial input to these tests in the form
of diffractive parton densities and of the generalised gluon distribution,
which are needed to calculate the hard part $\sigma_{\text{hard}}$ of the
cross section.  The importance of rescattering can also be probed in
diffractive photoproduction at HERA, given the hadronic component of a
real photon.  This has turned out to be more complicated than initially
thought, both from the experimental and the theoretical sides
\cite{diff-jet-talks,Klasen:2008ah}.  It is probably too early to draw
final conclusions here, but ultimately these studies might teach us more
about the double nature of the photon as a pointlike particle and a hadron
than about rescattering dynamics in diffraction.


\section{Saturation and the dipole formalism}

Parton saturation, caused by nonlinear dynamics that sets in when parton
densities become very large, has become a central topic in high-energy
QCD.  It allows us to study a field theory in a strongly coupled regime,
but with a small coupling constant, which makes it possible to use
perturbative methods.  Beyond this intrinsic interest, parton saturation
entails the breakdown of a description based on collinear factorisation
and DGLAP evolution.  To quantify nonlinear effects at small $x_B$ and
moderate $Q^2$ at HERA is hence relevant for assessing the limits of
precision in DGLAP based extractions of parton densities.  For generic
reasons, one expects such nonlinear, higher-twist corrections to be
stronger in $F_L$ than in~$F_2$.  Quantitative estimates based on the
colour dipole model have been given several years ago
\cite{Bartels:2000hv}, and work is underway to update these estimates
taking into account the progress in dipole phenomenology
\cite{Motyka-HERA-LHC}.

A convenient framework to describe parton saturation in $ep$ collisions is
the colour dipole formalism.  It permits the calculation of many inclusive
and diffractive processes---from the inclusive structure functions
$F_2^{}$, $F_2^{\text{charm}}$, $F_L^{}$ and their diffractive
counterparts to exclusive vector meson production and deeply virtual
Compton scattering---with the \emph{same} non-perturbative input, namely
the scattering amplitude for a dipole on a proton target.  The associated
phenomenology is very successful, as presented in \cite{Watt-talk} at this
meeting.  On the theoretical side, this formalism is, however, still
largely restricted to leading order in $\alpha_s$.  The fluctuation of a
$\gamma^*$ into $q\bar{q}$ is readily taken into account, but the next
highest Fock state $q\bar{q}g$ has so far only been taken in certain
approximations, which limits a reliable description of inclusive
diffraction.  The evolution of the dipole scattering amplitude with energy
is described by the BFKL or the Balitsky-Kovchegov equations, whose
leading-order solutions cannot account for the energy dependence seen in
experiment.  In practice one therefore typically takes a functional form
of the dipole scattering amplitude motivated by theory, but fits the
relevant parameters to data.  Note that this is different from DGLAP type
fits in collinear factorisation, where the relevant parton distributions
are parametrised at a starting scale but evolved using the perturbative
evolution equations.  It remains an outstanding task to formulate BFKL and
saturation dynamics in a dipole framework at NLO, in a manner that would
allow one to pursue phenomenological analyses.

To which extent nonlinear dynamics is seen in HERA data remains rather
controversial.  Current saturation models find that the virtualities where
saturation effects become important are below a GeV at HERA energies,
which limits the possibilities of controlled perturbative calculations.  A
prominent experimental observation is the very flat energy dependence for
the ratio $F_2^D /F_2^{}$ of diffractive and total structure functions at
given $Q^2$.  This is explained in a natural way by the saturation
mechanism \cite{GolecBiernat:1999qd}.  It must, however, be noted that
many dipole models, including versions without saturation, provide a good
description of $F_2^D$ and $F_2^{}$ in a wide kinematic range
\cite{Forshaw-talk}.  To better assess the situation, it may be helpful to
compare models with data specifically for the structure function ratio.

A striking feature observed in HERA small-$x$ data is geometric scaling,
i.e., the dependence of the total $\gamma^* p$ cross section (and also of
diffractive cross sections) on a single scaling variable $\tau(x_B, Q^2)$
\cite{Royon-talk}.  This is sometimes presented as evidence for saturation
dynamics.  However, several investigations have shown that, both
analytically and numerically, one finds approximate geometric scaling also
from the DGLAP and the BFKL equations \cite{Kwiecinski:2002ep}.  To infer
from geometric scaling on the underlying dynamics, one may have to focus
on the \emph{deviations} from exact scaling, which should differ among the
various dynamical mechanisms.  How well this can be quantified
theoretically, and whether the precision and kinematic lever arm of the
HERA data are sufficient for such a study, remains to be seen.

There are prospects to pursue the HERA studies of saturation effects for
the much smaller parton momentum fractions achievable at LHC.  This will
require detection of particles at very forward rapidities
\cite{Albrow:2000}.  Forward Drell-Yan pair production is of particular
interest from a theory point of view.  On the one hand, it permits a
description in the dipole picture \cite{Brodsky:1996nj} and can thus be
closely related with the studies performed at HERA.  On the other hand
this process is very well understood in collinear factorisation, with full
next-to-next-to-leading order results~\cite{Moch:2008dt} allowing for
precise ``non-saturated'' calculations.


\section*{Acknowledgments}

It is a pleasure to acknowledge discussions with D.Yu.~Ivanov and L.
Motyka, and to thank R.~Devenish, M.~Wing, and their co-organisers for
hosting a great conference.


\begin{footnotesize}

\end{footnotesize}


\end{document}